\pdfoutput=1

\documentclass[11pt]{article}

\usepackage{acl}
\usepackage{times}
\usepackage{latexsym}

\usepackage[T1]{fontenc}

\usepackage[utf8]{inputenc}

\usepackage{microtype}

\usepackage{inconsolata}

\usepackage[table]{xcolor}
\usepackage{stfloats}
\usepackage{graphicx}
\usepackage{balance}
\usepackage{multirow}
\usepackage{graphicx}
\usepackage{amssymb}
\usepackage{pifont}
\usepackage{tabularx}
\usepackage{multirow}
\usepackage{graphicx}
\usepackage{booktabs}   
\usepackage{tabularx}  
\usepackage{array}     
\usepackage{caption}   
\usepackage{float}
\usepackage{caption}
\usepackage{tcolorbox}
\usepackage{array}
\usepackage{subcaption}
\usepackage{arydshln} 
\usepackage{listings} 
\usepackage{courier}  
\usepackage{amsmath}  
\usepackage{anyfontsize}
\usepackage{array} 
\usepackage[loadshadowlibrary]{todonotes}
\usepackage{listings}
\usepackage{xcolor}
\usepackage{hyperref}
\usepackage{enumitem}
\usepackage{placeins}
\usepackage{graphicx}
\usepackage{balance}
\usepackage{amssymb}
\usepackage{pifont}
\usepackage{tikz}
\usetikzlibrary{arrows.meta, positioning, shapes, calc}
\usepackage{xcolor}
\usepackage{tabularx}
\usepackage{multirow}
\usepackage{booktabs}   
\usepackage{array}     
\usepackage{caption}  
\usepackage{float}
\usepackage{tcolorbox}
\usepackage{subcaption}
\usepackage{arydshln} 
\usepackage{courier}  
\usepackage{amsmath}  
\newcommand{\best}{\ding{72}}
\newcommand{\good}{\ding{118}}
\usepackage{anyfontsize}
\usepackage{listings}
\newcommand{\hlblue}[1]{\begingroup\setlength{\fboxsep}{1pt}\colorbox{cyan!20}{\strut #1}\endgroup}
\newcommand{\hlorange}[1]{\begingroup\setlength{\fboxsep}{1pt}\colorbox{orange!20}{\strut #1}\endgroup}
\newcommand{\hlgreen}[1]{\begingroup\setlength{\fboxsep}{1pt}\colorbox{green!10}{\strut #1}\endgroup}
\newcommand{\hlpink}[1]{\begingroup\setlength{\fboxsep}{1pt}\colorbox{red!10!white}{\strut #1}\endgroup}

\usepackage{hyperref}
\title{ReleaseEval: A Benchmark for Evaluating Language Models in Automated Release Note Generation}

\author{
  Qianru Meng \and Zhaochun Ren \and Joost Visser \\
  The Leiden Institute of Advanced Computer Science (LIACS), Leiden University \\
  \texttt{\{q.r.meng, z.ren, j.m.w.visser\}@liacs.leidenuniv.nl}
}

\begin{document}
\maketitle
\begin{abstract}

Automated release note generation addresses the challenge of documenting frequent software updates, where manual efforts are time-consuming and prone to human error. Although recent advances in language models further enhance this process, progress remains hindered by dataset limitations, including the lack of explicit licensing and limited reproducibility, and incomplete task design that relies mainly on commit messages for summarization while overlooking fine-grained contexts such as commit hierarchies and code changes. To fill this gap, we introduce \textsc{ReleaseEval}, a reproducible and openly licensed benchmark designed to systematically evaluate language models for automated release note generation. \textsc{ReleaseEval} comprises 94,987 release notes from 3,369 repositories across 6 programming languages, and supports three task settings with three levels of input granularity: (1) \textit{commit2sum}, which generates release notes from commit messages; (2) \textit{tree2sum}, which incorporates commit tree structures; and (3) \textit{diff2sum}, which leverages fine-grained code diffs. Both automated and human evaluations show that large language models consistently outperform traditional baselines across all tasks, achieving substantial gains on \textit{tree2sum}, while still struggling on \textit{diff2sum}. These findings highlight LLMs’ proficiency in leveraging structured information while revealing challenges in abstracting from long code diffs.



\end{abstract}

\section{Introduction}
\label{sec:introduction}


Release notes are key documents in software projects, summarizing changes, enhancements, and fixes across versions \cite{alali2008typical,maalej2010work,shihab2013mining,yu2009mining}. As systems evolve rapidly, manual documentation becomes costly and error-prone, motivating research on automated release note generation. Early work focused on extractive methods \cite{moratanch2017extractive}, while later studies adopted neural abstractive approaches \cite{jiang2021deeprelease,kamezawa2022rnsum}. More recently, large language models (LLMs) have further advanced this task by offering greater fluency and contextual understanding \cite{daneshyan2025smartnote}.

\begin{figure*}[!t]
    \centering
\includegraphics[width=\linewidth]{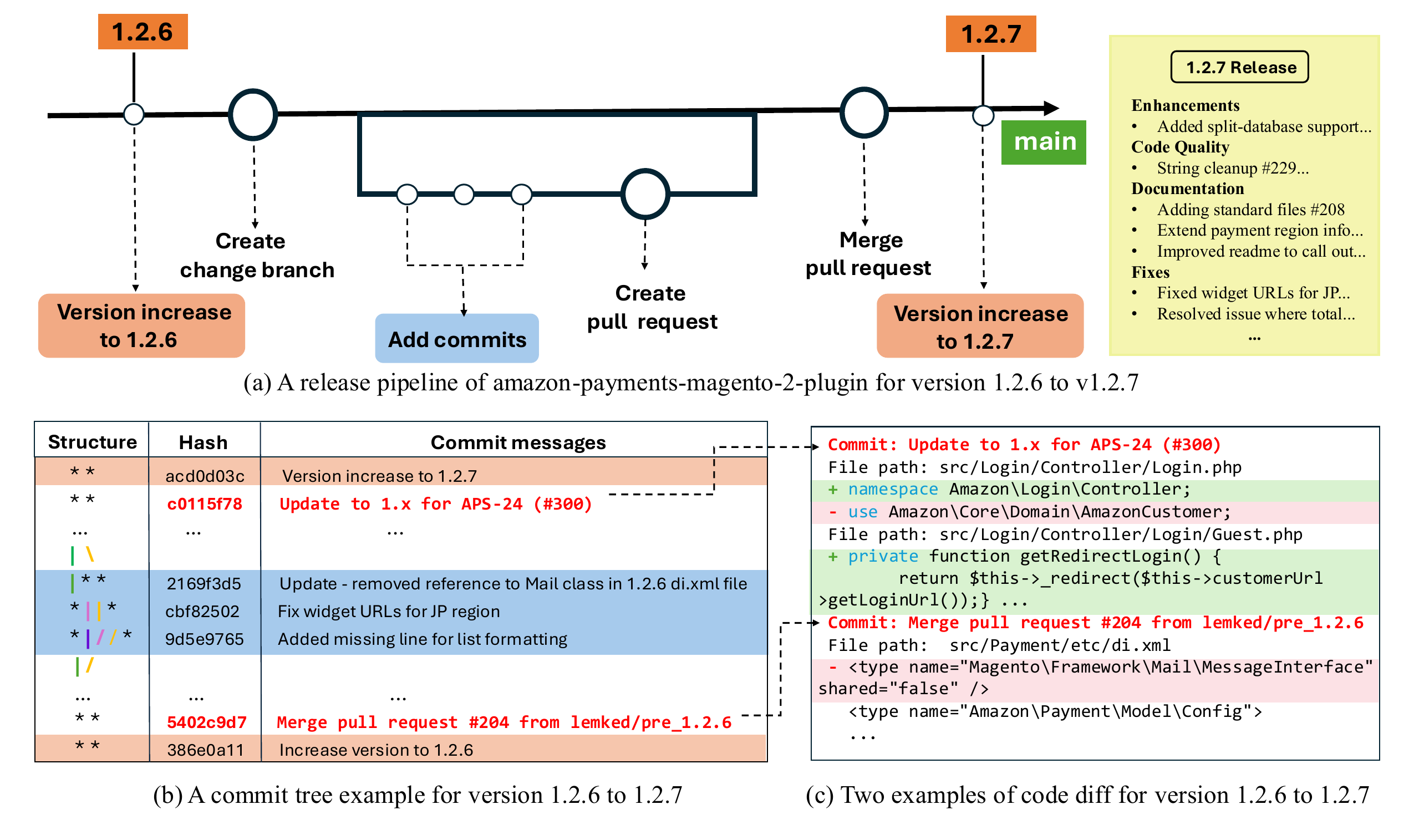}
    \caption{The details of commits from Amazon Payments Magento 2 plugin. Figure (a): a release pipeline which includes branch-level commits. Figure (b): commit tree, which includes commit messages and their structures. Figure (c): two fine-grained code diffs, corresponding to the messages. Note: (a) and (b) highlight matched commits: blue denotes commits added on the change branch, orange indicates version-bump commits on the main branch.
In (c), red marks the commits corresponding to (b), green highlights added code, and pink marks deleted code.} 
    
    \label{fig:c_pr_rn}
\end{figure*}

Despite this progress, existing benchmark datasets \cite{nath2022towards,jiang2021deeprelease,kamezawa2022rnsum,daneshyan2025smartnote} for release note generation suffer from two major shortcomings that limit their research and practical value. First, licensing and reproducibility remain unresolved. Although many are open-sourced, the absence of explicit distribution licenses prevents their legal reuse and redistribution. Moreover, some datasets have limited reproducibility. Even when the repositories used in data collection are documented, these repositories evolve over time, leaving parts of the datasets unavailable \cite{kamezawa2022rnsum}. Second, dataset design directly affects the quality of release note generation. Most existing datasets define the task narrowly as commit-to-release generation, thereby overlooking critical information in the structural hierarchy of commits and fine-grained code changes. As illustrated in Figure~\ref{fig:c_pr_rn}(b), software commits typically contain not only commit messages but also a branching structure with parent–child relationships, which we refer to as the commit tree. Furthermore, as shown in Figure~\ref{fig:c_pr_rn}(c), code diffs provide precise, line-level details essential for understanding the nature and impact of changes.

To address these gaps, we introduce \textsc{ReleaseEval}, an open-access and reproducible benchmark, designed to assess current models for three release note generation tasks with varying input granularity. It extends the traditional task \textit{commit2sum} by introducing two new tasks: \textit{tree2sum}, which incorporates structural commit hierarchies, and \textit{diff2sum}, which leverages line-level code changes. \textsc{ReleaseEval} contains 94,987 release notes from 3,369 repositories across six programming languages. By conducting both automatic and manual evaluations across nine language models, we observe findings from two perspectives: (1) Model performance: LLMs consistently outperform existing pre-training language models across all three tasks. Among them, fine-tuned Ministral-8B achieves the best performance on \textit{tree2sum}, with 42.77\% BLEU-4, 53.14\% ROUGE-L, and 55.79\% METEOR. (2) Task design: \textit{tree2sum} consistently achieves the best performance for the same model, benefiting from additional structural information provided by the commit tree. In contrast, fine-grained code diffs introduce substantial noise and significantly increase the input context, which limits the generation performance in the \textit{diff2sum} task. Our contributions are summarized as follows:
\begin{itemize}
    \item We present \textsc{ReleaseEval}, a large-scale, fully reproducible benchmark with open-source licensing for release note generation, covering 94,987 release notes in 3,369 projects across six programming languages.
    
    \item We propose two novel tasks in \textsc{ReleaseEval}, \textit{tree2sum} and \textit{diff2sum}, which incorporate fine-grained structural and code-level information, respectively.

    \item We evaluate seven open-source LLMs across three tasks, revealing their capabilities across different model scales and input granularities.
\end{itemize}


\begin{table*}[htbp]
\centering
\small
\begin{tabularx}{\textwidth}{l *{5}{>{\centering\arraybackslash}X}}

\toprule
\textbf{Features} & \textbf{Nath \cite{nath2022towards}} & \textbf{DeepRelease \cite{jiang2021deeprelease}} & \textbf{RNSum \cite{kamezawa2022rnsum}} & \textbf{SmartNote \cite{daneshyan2025smartnote}} & \textbf{ReleaseEval} \\ 
\midrule
Release Notes & 1,213 & 37,000 & 81,996 & 21,882 & 94,987 \\ 
Repositories & 13 & 561 & 7,216 & 272 & 3,369 \\ 
Reproducible & \checkmark & \checkmark & Semi & \checkmark & \checkmark \\ 
Deduplicated & \checkmark & \checkmark & \checkmark & \checkmark & \checkmark \\ 
License-Aware Data Collection & \ding{55} & \ding{55} & \ding{55} & \ding{55} & \checkmark \\ 
Additional Filtering & \checkmark & \checkmark & \checkmark & \checkmark & \checkmark \\ 
Programming Languages & 3 & 8 & NA & 7 & 6 \\
\bottomrule
\end{tabularx}
\caption{Comparison between \textsc{ReleaseEval} and existing datasets for release note generation.}
\label{tab:data_compar}
\end{table*}

\section{Related Work}
\label{sec:related work}

\subsection{Approaches of Automated Release Note Generation} 

Prior work on automated release note generation can be categorized into extractive summarization and abstractive summarization methods.

Extractive summarization uses importance scores to extract key sentences or segments from sources (e.g., Pull Requests or change logs) and assembles them into summaries. Early rule-based methods include the semi-automated approach proposed by \citeauthor{klepper2016semi}~(\citeyear{klepper2016semi}) and the fully automated ARENA tool by \citeauthor{moreno2016arena}~(\citeyear{moreno2016arena})
. Recent work leverages embedding-based ranking to capture semantic representations. For example, \citeauthor{nath2022towards}~(\citeyear{nath2022towards})
combine GloVe embeddings \cite{pennington2014glove} with the TextRank algorithm \cite{mihalcea2004textrank} to select important commits as release notes.

Abstractive summarization methods for release note generation often follow the Classwise Abstractive Summarization paradigm \cite{kamezawa2022rnsum}, where commit messages are first classified before generating release notes. Jiang et al. \cite{jiang2021deeprelease} propose DeepRelease, which employs an LSTM-based encoder–decoder to generate release notes from Pull Requests (PRs), using FastText for PRs classification. Similarly, \citeauthor{klepper2016semi}~(\citeyear{klepper2016semi}) present a transformer-based framework that leverages BERT for classifying commit messages and BART \cite{lewis2019bart} for generating release notes. Daneshyan et al. \cite{daneshyan2025smartnote} introduce SmartNote, which combines machine learning classifiers for commit classification with large language models for release note generation.

\subsection{Datasets of Automated Release Note Generation} 

Despite advancements in models for release note generation, the creation and selection of appropriate datasets remain a key challenge, which is the main focus of this paper. Researchers have developed various datasets for evaluating existing methods. Nath et al.~\cite{nath2022towards} built a dataset of 1,200 GitHub releases paired with commit messages and PR titles to evaluate their extractive method. To evaluate DeepRelease, Jiang et al.~\cite{jiang2021deeprelease} developed a dataset of over 46K release notes associated with PRs. Kamezawa et al.~\cite{kamezawa2022rnsum} introduced RNSum, a dataset consisting of 82,000 release notes derived from commit messages, and used it to evaluate multiple abstractive summarization methods. Recently, Daneshyan et al.~\cite{daneshyan2025smartnote} released a dataset of over 21,000 release notes aligned with commit messages to evaluate their LLM-based method, SmartNote. As shown in Table~\ref{tab:data_compar}, existing datasets each exhibit one or more limitations, such as limited reproducibility (e.g., RNSum) and incomplete licensing information (e.g., Nath, DeepRelease, RNSum, and SmartNote). Furthermore, these datasets constrain the input to commit messages, which omits the value of fine-grained context. These limitations highlight the importance of developing a benchmark with broader coverage.

\section{ReleaseEval}
\label{sec:framework}

\subsection{Overall Framework}

The construction of \textsc{ReleaseEval} follows a structured framework encompassing data collection and processing, benchmark construction and evaluation, as illustrated in Figure~\ref{fig:pipeline}. The following section provides a detailed explanation of each component.

\subsection{Data Collection and Processing}
\label{sec:dataset}

\subsubsection{Project Selection}

The projects used in our dataset are sourced from CommitChronicle~\cite{eliseeva2023commit_chronicle}, a benchmark for commit message generation that provides openly licensed data from over 12,000 repositories. To ensure project quality and activity, we first filter for repositories with more than 50 stars and at least 10 contributors. However, as ~\citeauthor{hu2016influence} (\citeyear{hu2016influence}) point out, starred projects do not necessarily represent the most important or actively maintained ones. Therefore, we further refine our selection by retaining only repositories with an issue resolution rate above 40\% and with recent commit activity between August 1, 2017, and October 5, 2024. To ensure diversity, we also analyze repository ownership and find that 86\% belong to organizations while 14\% are maintained by individual developers. Additionally, we focus on repositories implemented in six of the most widely used programming languages on GitHub—TypeScript, JavaScript, Go, Python, Java, and PHP—selected from an initial pool of 20 programming languages. After applying all filters, we obtain a final selection of 3,369 projects from the original 12.4K repositories.

\begin{figure*}
  \centering
  \includegraphics[width=\linewidth]{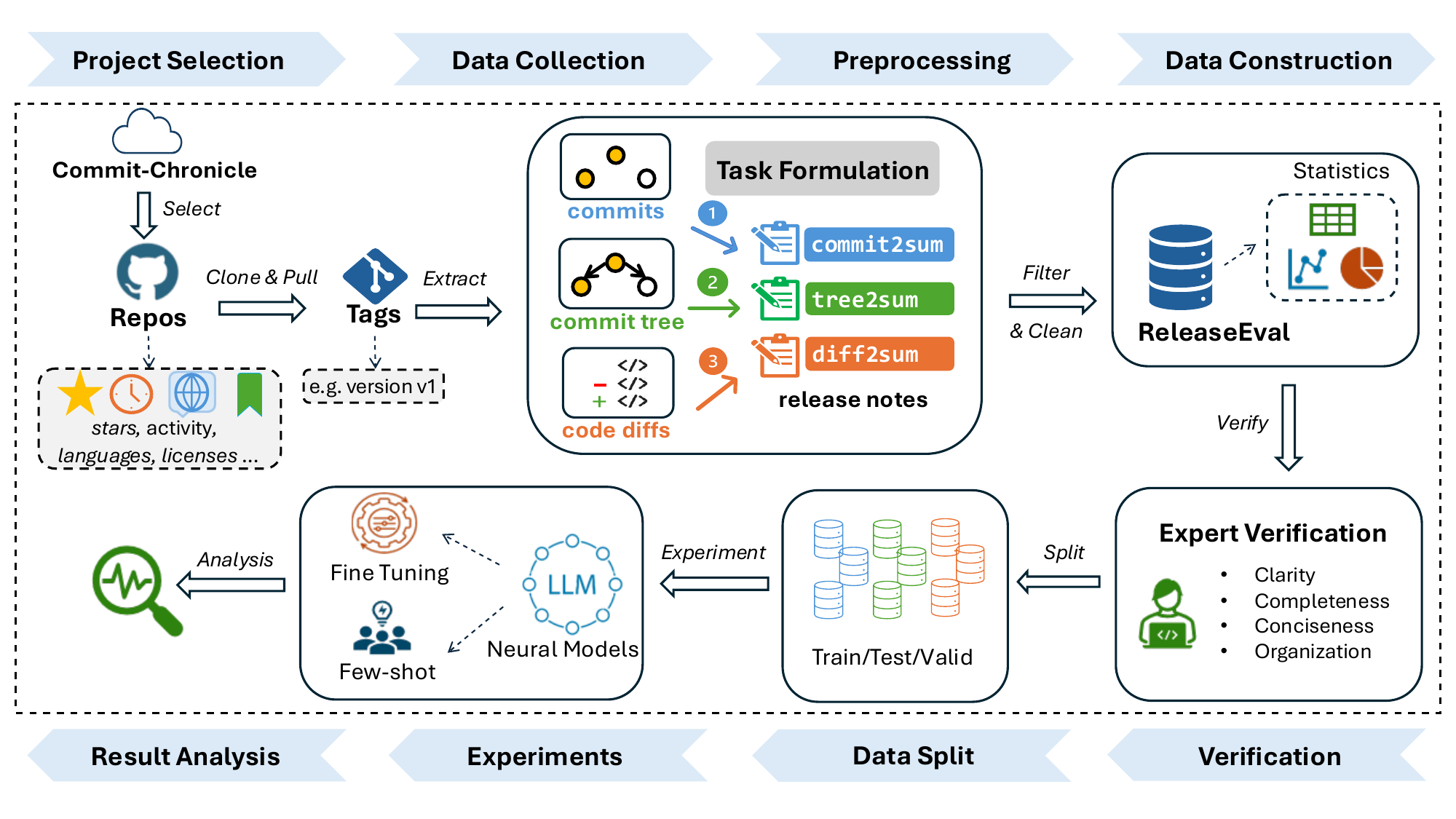}
\caption{The overall framework of \textsc{ReleaseEval}: (1) selecting projects from the CommitChronicle dataset \cite{eliseeva2023commit_chronicle}; (2) collecting raw data of releases, commits, and code diffs based on the selected projects; (3) preprocessing the data using filtering strategies; (4) generating dataset statistics; (5) verifying dataset quality through expert evaluation; (6) splitting the dataset for training and evaluation; (7) conducting experiments; and (8) analyzing and discussing the results.}
  \label{fig:pipeline}  
\end{figure*}

\subsubsection{Dataset Collection}
After selecting the target projects, we systematically collected a comprehensive set of attributes, summarized in Table~\ref{tab:attributes} (see in Appendix \ref{app:dataset_detail}), that capture the metadata associated with releases, commits, and code changes. To ensure scalability and reproducibility, we implemented an automated data collection pipeline leveraging the GitHub API~\cite{github_api} for retrieving release notes and commit trees, and employed PyDriller~\cite{spadini2018PyDriller} to extract commit metadata and fine-grained code diffs. 

\subsubsection{Dataset Processing} To obtain high-quality data, we apply several filtering strategies targeting the following attributes.


\textbf{Versions.} The first step is to extract and clean repository versions, enabling mapping of release notes to corresponding commits. For example, the release notes for version 1.1.1 are based on the commits between the previous version (e.g., 1.1.0) and the current version (1.1.1). Cleaning versions helps filter out pre-release or test versions (e.g., "beta," "alpha," or "rc"). Similar to filtering strategy used by \cite{kamezawa2022rnsum}, we use regular expressions to identify and remove invalid versions, followed by validating core version numbers using the packaging.version library. After filtering, the valid versions are sorted chronologically, and only consecutive version pairs are retained.

\textbf{Release Notes.}
Based on the cleaned versions, we retrieve the corresponding release notes using the GitHub API. Based on previous work \cite{nath2022towards,kamezawa2022rnsum,daneshyan2025smartnote}, we choose the following three subsequent filtering strategies to clean the release notes. (1) \textit{Coarse Filtering:} We first apply general filtering criteria to remove invalid release notes, including those that are empty, consist of only one sentence, are duplicates, or contain fewer than 10 tokens. We also discard overly long release notes to save computational resources \cite{kamezawa2022rnsum}. These cases represent less than 3\% of the dataset and have negligible effect on data coverage and representativeness. (2) \textit{Category-based Filtering:} To ensure semantic coverage and informativeness, we adopt the category-matching tool proposed by \citeauthor{kamezawa2022rnsum}~(\citeyear{kamezawa2022rnsum}) to exclude release notes that cover fewer than two of the following categories: features, improvements, bug fixes, and deprecations. (3) \textit{Noise Filtering:} For the remaining release notes, we remove token-level noise such as issue versions, emojis, and SHA identifiers, following prior commit filtering studies \cite{jiang2017automatically,xu2019commit,wang2021context,shi2022race,dong2022fira}.

\textbf{Commits.} Using the retained consecutive version pairs, we employ PyDriller to collect commit messages, including the corresponding commit hash, author, timestamp, and code diffs. We apply the same \textit{noise filtering} strategy to commit messages as to release notes.

\textbf{Commit Trees.} Similarly, using consecutive version pairs, we access commit tree information through the GitHub API, applying the same filtering strategy as we use for commit messages to remove irrelevant information. The commit tree is stored and presented in a plain text format, using the symbols specified in Table \ref{tab:tree_structure} (see in Appendix \ref{app:dataset_detail}) to represent the structure.

\textbf{Code Diffs.} We construct diffs at the version level by combining file-level modifications with the corresponding code changes, where each modification records the file path, change type (e.g., modify, add and delete), and related code changes.

\subsubsection{Dataset statistics}

\begin{figure}[!t]
  \centering
  \includegraphics[width=0.75\linewidth]{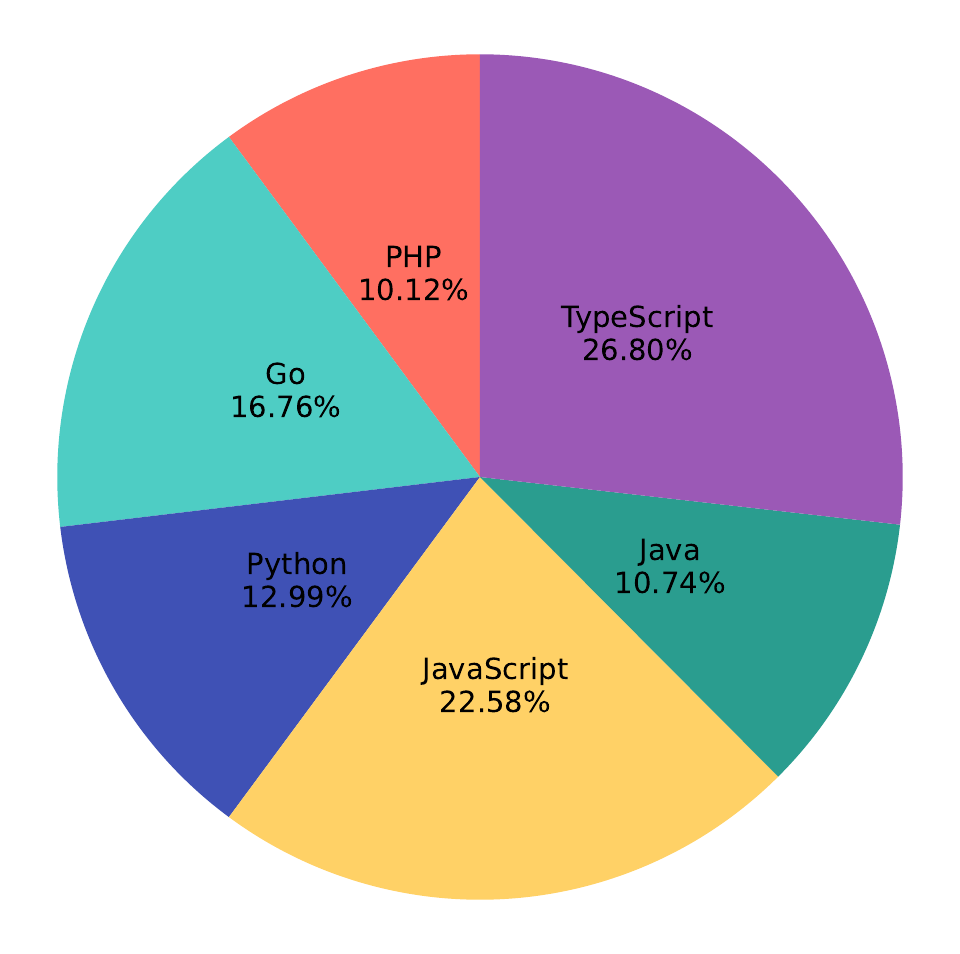}
  \caption{Programming Language Distribution of \textsc{ReleaseEval}.}
  \label{fig:lan_distribution}
\end{figure}

After processing, Table \ref{tab:repo_data} shows that the dataset includes 3,369 repositories, resulting in 94,987 release notes and commit trees, alongside 1,074,577 commit messages and code diffs. Based on the average length of key attributes, the code diffs are significantly longer than other input types, with an average length exceeding 1,300 tokens. Figure~\ref{fig:lan_distribution} illustrates the distribution of programming languages in \textsc{ReleaseEval}, with TypeScript and JavaScript being the most dominant.

\subsubsection{Quality verification}
\label{sec:data-verification}
Finally, we conduct an expert verification to ensure the data quality. Following Daneshyan et al. \cite{daneshyan2025smartnote}, we used four metrics rated on a 5-point Likert scale for evaluation. (1) \textit{Completeness}: Ensures the release note includes every significant change made between versions. (2) \textit{Clarity}: Verifies that each update is clearly and correctly presented. (3) \textit{Conciseness}: Confirms the note delivers only key changes without unnecessary details. (4) \textit{Organisation}: Evaluates whether the content is arranged in a clear and logical structure. We invited three experts (each with 6–8 years of programming experience) to assess 1,000 release notes randomly selected from 96 projects. To assess inter-rater reliability, we measured overall inter-annotator agreement, which reached 0.85 (Fleiss’ Kappa \cite{fleiss1971mns}). 

\begin{table}
\centering
\small
\begin{tabularx}{\linewidth}{@{}Xl@{}}
\toprule
\textbf{Feature} & \textbf{Total} \\
\midrule
Repositories                      & 3,369     \\
Release Notes                     & 94,987    \\
Commit Trees                      & 94,987    \\
Commit Messages                   & 1,074,577 \\
Code diffs                  & 1,074,577 \\
Avg.\ \# Commits per Release Note & 14        \\
Avg.\ \# Tokens per Release Notes & 198       \\
Avg.\ \# Tokens per Commit Message & 14        \\
Avg.\ \# Tokens per Commit Tree   & 144       \\
Avg.\ \# Tokens per Code Difference & 1,348     \\
\bottomrule
\end{tabularx}
\caption{Dataset Statistics of \textsc{ReleaseEval}}
\label{tab:repo_data}
\end{table}

Expert review confirms the high quality of the benchmark, with evaluation results included in the released repository. Among all metrics, organisation and conciseness received the highest scores (4.74 and 4.33, respectively). Notably, over 95\% of release notes in the dataset follow a standard structure with clearly defined categories. Clarity achieved a moderate score (4.08), while completeness scored the lowest (3.85), which aligns with prior findings \cite{wu2022demystifying} that missing or incorrect change information is a common issue in release note practices. The relatively lower clarity scores observed in 68\% of small-scale projects (e.g., storybook, vaadin, etc.) may be due to their direct reuse of raw commit messages without paraphrasing. In contrast, large-scale projects (e.g., Google, Kubernetes, etc.) tend to produce high-quality release notes across all four evaluation metrics. Despite variations in writing style, these notes generally follow a standard structure and present information clearly and coherently.

\subsection{Task Formulation} 

As illustrated in Figure~\ref{fig:pipeline}, we define three tasks on \textsc{ReleaseEval}. The first task, commit2sum, aims to generate summaries from a list of commit messages. The second task, \textit{tree2sum}, utilizes structured commit tree as input, which includes both parent-child dependencies and commit messages. The symbols used to construct the commit tree are detailed in Appendix~\ref{tab:tree_structure}. The third task, \textit{diff2sum}, leverages the fine-grained code diffs between software versions.

\subsection{Data Split}
We split the dataset into training, validation, and test sets using an 8:1:1 ratio, applied consistently across all programming languages. This yields 75,993 data for training, 9,495 for validation, and 9,499 for testing.

\section{Experiments}
\label{sec:experiment}








\subsection{Model Selection}
To comprehensively assess the effectiveness of representative language models on \textsc{ReleaseEval}, we evaluate seven models, covering both encoder–decoder and decoder-only architectures. Specifically, we select BART~\cite{lewis2019bart}, a strong baseline previously applied to release note generation~\cite{kamezawa2022rnsum}, and T5~\cite{raffel2020t5}, a widely used encoder–decoder model for summarization and other text generation tasks. For decoder-only LLMs, we include a set of open-source models with different scales and architectures: LLaMA3.1 (8B) and LLaMA3.3 (70B) \cite{touvron2024llama3}, Mistral (8B and 22B) \cite{jiang2023mistral}, and Qwen2.5 (7B, 32B, and 72B) \cite{baichuan2024qwen2}.

\subsection{Evaluation Metrics}
To automatically evaluate the quality of the generated release notes, we use the widely adopted metrics BLEU-4 \cite{papineni2002bleu}, ROUGE-L \cite{lin2004rouge}, and METEOR \cite{banerjee2005meteor}, which are commonly used in prior work \cite{nath2022towards,kamezawa2022rnsum,jiang2021deeprelease} to capture lexical overlap and semantic similarity. For manual evaluation, we assess the generated release notes using four dimensions: Completeness, Clarity, Conciseness, and Organisation, as previously introduced in Section~\ref{sec:data-verification}.

\subsection{Implementation Details} 

We evaluate the models using both fine-tuning \cite{howard2018universal} and few-shot settings \cite{brown2020language}. Specifically, due to limited computational resources, we fine-tune three smaller LLMs in the 7B–8B range (Qwen2.5-7B, LLaMA3.1-8B, and Mistral-8B), along with BART and T5. For few-shot evaluation, we apply zero-shot, two-shot, and four-shot prompting to all LLMs. In the zero-shot setting, all seven models are evaluated across all three tasks. For the two-shot and four-shot settings, we focus on the \textit{commit2sum} and \textit{tree2sum} tasks, excluding \textit{diff2sum} due to its excessive input length in the few-shot setting. Table~\ref{tab:combined_config} summarizes the hyperparameter settings and hardware configurations, and Table~\ref{tab:combinedprompts} provides the prompt templates for the few-shot setting; both tables are included in Appendix~\ref{app:exp_detail}.

\setlength{\tabcolsep}{3pt}
\begin{table*}[htbp]
\centering
\small
\begin{tabularx}{\textwidth}{l *{9}{>{\centering\arraybackslash}X}}
\toprule
\multirow{2}{*}{Method}
 & \multicolumn{3}{c}{\textbf{commit2sum}}  & \multicolumn{3}{c}{\textbf{tree2sum}}  & \multicolumn{3}{c}{\textbf{diff2sum}} \\
\cmidrule(lr){2-4} \cmidrule(lr){5-7} \cmidrule(lr){8-10}
 & BLEU-4 & \mbox{ROUGE-L} & METEOR  & BLEU-4 & \mbox{ROUGE-L} & METEOR  & BLEU-4 & \mbox{ROUGE-L} & METEOR \\
\midrule
BART     & 19.96 & 32.48 & 32.23  & 25.36 & 37.23 & 37.65  & 17.52 & 28.11 & 24.38 \\
T5       & 22.43 & 34.02 & 34.13  & 27.61 & 37.75 & 38.83  & 18.20 & 26.37 & 24.29 \\
Qwen-2.5-7B     & 36.42 & 48.39 & 50.51  & 41.19 & 51.47 & 54.05  & 19.00 & 28.11 & 36.94 \\
LLaMA3.1-8B \best & \textbf{37.79} & \textbf{49.90} & \textbf{52.11}  & 42.64 & 52.91 & 55.37  & \textbf{33.91}  & \textbf{44.81} & \textbf{45.37} \\
Ministral-8B \good & 37.42 & 49.74 & 51.92  & \textbf{42.77} & \textbf{53.14} & \textbf{55.79}  & 32.79 & 43.57 & 44.82 \\
\bottomrule
\end{tabularx}
\caption{Fine-tuning performance (\%) of various language models on three tasks. Best performance is marked bold.}
\label{tab:overall_results}
\end{table*}
\begin{table*}
  \centering
  \small
  \begin{tabularx}{\textwidth}{l *{9}{>{\centering\arraybackslash}X}}
    \toprule
    Model 
      & \multicolumn{3}{c}{\textbf{commit2sum}}
      & \multicolumn{3}{c}{\textbf{tree2sum}}
      & \multicolumn{3}{c}{\textbf{diff2sum}} \\
    \cmidrule(lr){2-4} \cmidrule(lr){5-7} \cmidrule(lr){8-10}
      & BLEU-4 & \mbox{ROUGE-L} & METEOR  
      & BLEU-4 & \mbox{ROUGE-L} & METEOR  
      & BLEU-4 & \mbox{ROUGE-L} & METEOR \\
    \midrule
    \cellcolor[gray]{0.9}{\textbf{7–8B Models}} & & & & & & & & & \\
    Qwen2.5-7B          &  3.22 & 25.67 & 23.83  &  4.70 & 26.31 & 26.03  &  3.05 & 21.78 & 20.48 \\
    LLaMA3.1-8B \best   &  \textbf{6.07} & 29.85 & \textbf{27.23}  &  \textbf{7.54} & 29.73 & \textbf{29.49}  &  5.12 & 25.78 & 21.04 \\
    Ministral-8B \good  &  5.39 & \textbf{31.43} & 25.94  &  7.05 & \textbf{33.18} & 27.48  &  \textbf{5.29} & \textbf{30.02} & \textbf{23.79} \\
    \midrule
    \cellcolor[gray]{0.9}{\textbf{22–32B Models}} & & & & & & & & & \\
    Mistral-22B         &  4.46 & 28.35 & 25.79  &  5.70 & 30.17 & 26.09  &  4.37 & 26.71 & 24.83 \\
    Qwen2.5-32B         &  3.14 & 26.54 & 24.29  &  4.69 & 26.88 & 26.17  &  3.52 & 27.71 & 24.70 \\
    \midrule
    \cellcolor[gray]{0.9}{\textbf{70–72B Models}} & & & & & & & & & \\
    Llama3.3-70B        &  5.65 & 27.79 & 26.31  &  7.49 & 30.27 & 29.46  &  5.21 & 27.00 & 25.94 \\
    Qwen2.5-72B         &  3.43 & 27.40 & 24.31  &  3.35 & 26.80 & 23.19  &  3.38 & 26.32 & 23.12 \\
    \bottomrule
  \end{tabularx}
  \caption{Zero‐shot performance (\%) of LLMs at different scales on three tasks. Best performance is marked bold.}
  \label{tab:zero_shot_all}
\end{table*}

\section{Results and Analysis}
\label{sec:results}


\subsection{Performance of Fine-tuned LMs}

\textbf{Fine-Tuning:} The results of fine-tuning language models on \textsc{ReleaseEval} across three tasks are presented in Table~\ref{tab:overall_results}. As we can see, LLaMA3.1-8B achieves the best overall performance on \textit{commit2sum} and \textit{diff2sum}, while Ministral-8B performs best on \textit{tree2sum}. Across all tasks, three LLMs show substantial improvements over BART and T5, with the largest gains observed on \textit{diff2sum}—up to 15.7\% BLEU-4, 18.4\% ROUGE-L, and 21.1\% METEOR. This advantage of LLMs over BART and T5 can be attributed to their larger parameter scales and more diverse pre-training corpora, which enable better context understanding and accurate summarization.

\subsection{Performance of Few-shot LLMs}
\textbf{Zero-shot:} Table~\ref{tab:zero_shot_all} presents the zero-shot performance of LLMs across three tasks. LLaMA3.1-8B and Mistral-8B achieve the highest scores in this setting, particularly on \textit{tree2sum} (7.54\% BLEU-4, 33.18\% ROUGE-L, 29.49\% METEOR). However, performance in the zero-shot setting remains significantly lower than that of fine-tuned models, especially in BLEU-4, indicating that models capture high-level semantics but lack lexical fidelity. \textbf{Few-shot:} Figure~\ref{fig:few_shot_graph} presents LLM performance under few-shot settings on \textit{commit2sum} and \textit{tree2sum}. Most models benefit from more shots, with Mistral-22B achieving the best performance on both tasks. However, several models show a non-monotonic trend, with 2-shot performance dropping slightly before improving at 4-shot—consistent with prior findings that limited and non-diverse examples can bias the model and temporarily reduce generalization~\cite{min2022rethinking}.



\subsection{Impact of LLM Scale}
\begin{figure*}[!t]
    \centering
\includegraphics[width=\linewidth]{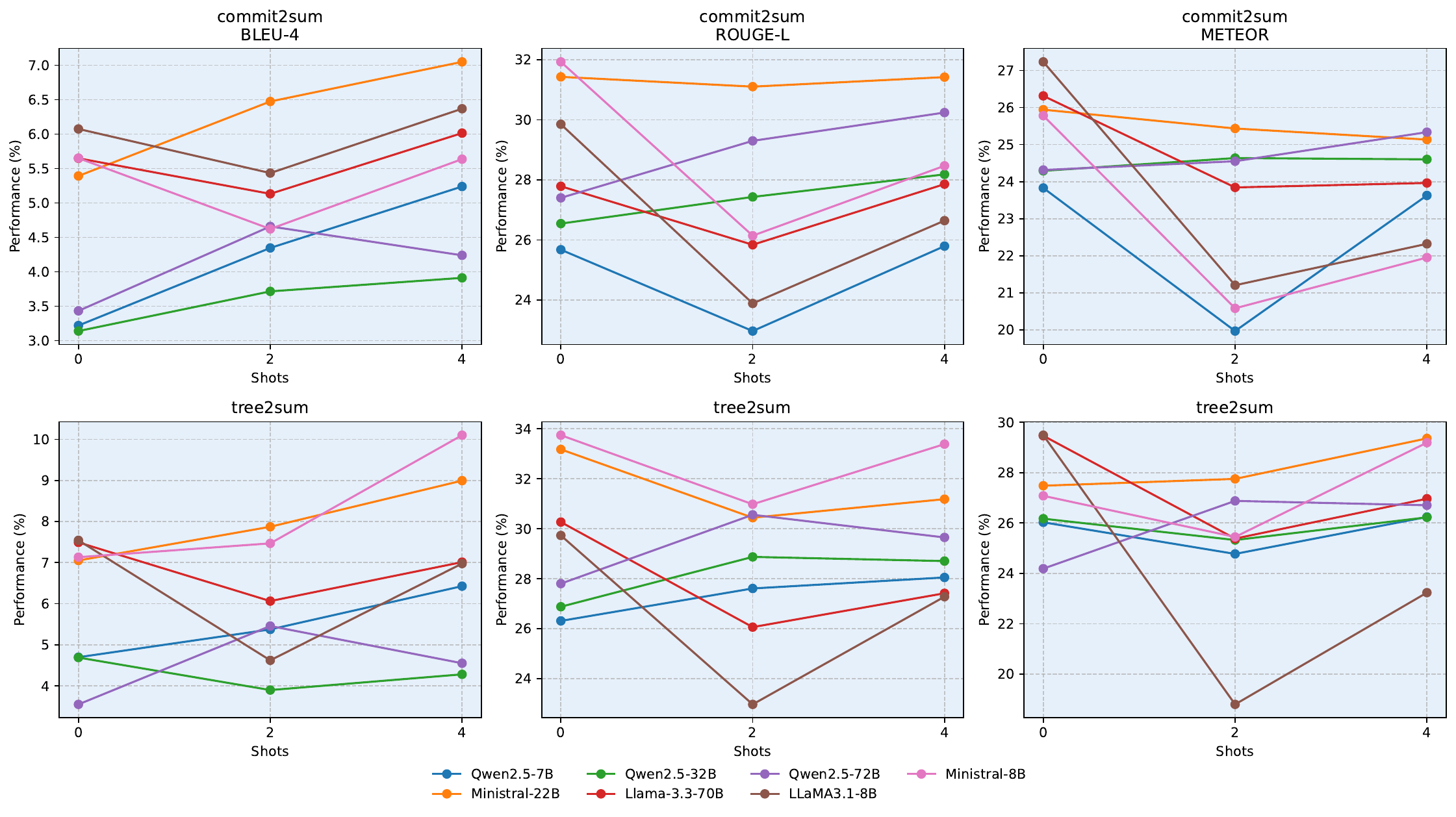}
    \caption{Few‐shot performance (\%) of LLMs at different scales on the \textit{commit2sum} and \textit{tree2sum} tasks across zero-, two-, and four-shot settings.}
    \label{fig:few_shot_graph}
\end{figure*}

As shown in Figure~\ref{fig:few_shot_graph}, smaller models such as LLaMA3.1-8B and Qwen2.5-7B exhibit noticeable fluctuations in the few-shot setting. In contrast, larger models (e.g., Qwen2.5-72B, LLaMA3.3-70B) tend to perform more consistently with shots increasing, likely due to their higher parameter capacity and greater robustness to longer inputs.


\subsection{Impact of inputs on Release Note Quality}
\label{sec:RQ2}

According to the results presented in Table~\ref{tab:overall_results}, Table~\ref{tab:zero_shot_all}, and Figure~\ref{fig:few_shot_graph}, LLMs consistently achieve the best performance on the \textit{tree2sum} task in both fine-tuning and few-shot settings, outperforming their results on \textit{commit2sum} and \textit{diff2sum}. This suggests that the additional structural context provided by \textit{tree2sum} enables models to generate more coherent and informative release notes. In contrast, the relatively lower performance on \textit{diff2sum} highlights the difficulty of generating high-quality summaries from raw code diffs, which are typically lengthy and low-level. To further investigate the influence of input types on the quality of generated outputs, we conducted a human evaluation based on a case study and quantitative analysis. 

\textbf{Case study:} 
We present a representative example from fine-tuned Ministral-8B in Table~\ref{tab:compare_3_task_example} (see Appendix \ref{app:case_study}), which shows the generated release notes for all three tasks alongside the corresponding reference note. Among the three inputs, \textit{tree2sum} achieves the highest BLEU-4 score (66.28), demonstrating strong alignment with the reference and capturing nearly all critical changes. \textit{commit2sum} also identifies several key updates (BLEU-4: 39.95), though it introduces some unrelated information. \textit{diff2sum}, by contrast, achieves the lowest score (BLEU-4: 16.10), missing several important updates and reflecting the difficulty of abstracting accurate summaries from raw code diffs. Despite differences in quality, all three outputs follow a consistent release note format, indicating that the model has effectively learned the standard structure and writing style during fine-tuning. To complement these insights, we conducted a quantitative evaluation with domain experts to assess the effect of input types on generation quality.

\begin{table}
\centering
\small
\begin{tabularx}{\linewidth}{
  @{}%
  l
  >{\centering\arraybackslash}p{1.3cm}   
  >{\centering\arraybackslash}p{1.2cm}   
  >{\centering\arraybackslash}p{1.3cm}   
  >{\centering\arraybackslash}p{1.3cm}   
  @{}
}
\toprule
\textbf{Data}       & \textbf{Comp.} & \textbf{Clar.} & \textbf{Conc.}  & \textbf{Org.}\\ 
\midrule
commit2sum          & 3.83           & 3.59           & 3.45            & 4.23        \\
tree2sum            & 4.02           & 3.68           & 3.47            & 4.23        \\
diff2sum            & 3.65           & 2.79           & 3.06            & 4.18        \\
\bottomrule
\end{tabularx}
\caption{Human evaluation results of release note quality across three tasks based on completeness (Comp.), clarity (Clar.), conciseness (Conc.) and organisation (Org.).}
\label{tab:human_eval}
\end{table}

\textbf{Quantitative Evaluation:} Following the evaluation protocol introduced in Section~\ref{sec:data-verification}, three experts evaluated 100 randomly selected outputs from fine-tuned Ministral-8B using four metrics: completeness, clarity, conciseness, and organisation. The evaluation yielded Fleiss’ kappa scores ranging from 0.82 to 0.89, indicating strong inter-rater agreement. The results are displayed in Table~\ref{tab:human_eval}, among the three tasks, \textit{tree2sum} achieves the highest overall quality, especially in completeness (4.02) and clarity (3.68), and performs comparably to \textit{commit2sum} in terms of conciseness and organisation. These results are consistent with our earlier findings, confirming the advantage of structural input for release note generation.

Interestingly, \textit{diff2sum} achieves a relatively high score in completeness (3.65), indicating that code diffs provide rich information. However, outputs based on code diffs are frequently redundant and repetitive, resulting in lower scores for clarity (2.79) and conciseness (3.06). This highlights a trade-off between content coverage and readability when using fine-grained but long inputs like code diffs. These results also demonstrate that effectively identifying key changes from rich but lengthy code diffs remains a major challenge.

\section{Conclusion}
\label{sec:conclusion}

In this paper, we introduce \textsc{ReleaseEval}, an open-source and reproducible benchmark for evaluating language models on three release note generation tasks: \textit{commit2sum}, \textit{tree2sum}, and \textit{diff2sum}. Our comprehensive evaluation reveals that LLMs achieve their best performance on \textit{tree2sum}, highlighting the advantage of structured commit trees in generating accurate and coherent release notes. In contrast, \textit{diff2sum} involves unstructured and lengthy code changes that reduce clarity and conciseness of the generated release notes, indicating that LLMs remain limited in abstracting semantic changes from long and complex code diffs. Therefore, \textsc{ReleaseEval} establishes itself as a robust benchmark for both practitioners and researchers, facilitating systematic evaluation and the development of LLMs for release note generation.

\section{Limitations}
\label{sec:limitations}

One limitation of our dataset concerns the nature of the ground-truth release notes. Prior work has shown that a clear and consistent style is important for readability~\cite{wu2022demystifying}, yet most of our annotated summaries follow a highly technical style that emphasizes low-level implementation details. While this style is appropriate for developers, it lacks broader contextual information, which reduces readability for non-technical stakeholders and limits the benchmark’s suitability for evaluating release notes aimed at diverse audiences. Another limitation lies in the scope of the \textit{tree2sum} and \textit{diff2sum} tasks. The \textit{diff2sum} setting often involves very long code diffs that can exceed model input length limits, making it difficult to preserve important details. Although techniques such as input truncation~\cite{liu2019text} and hierarchical encoding~\cite{rae2019compressive} have been proposed, effectively applying them in this domain remains challenging. The \textit{tree2sum} setting introduces additional complexity due to the hierarchical structure of commit trees and their dependencies, which may be better represented with graph-based methods such as Graph Neural Networks (GNNs) \cite{scarselli2008gnn}.

\bibstyle{plain}
\bibliography{latex/IEEEexample}

\begin{thebibliography}{38}
\providecommand{\natexlab}[1]{#1}

\bibitem[{Alali et~al.(2008)Alali, Kagdi, and Maletic}]{alali2008typical}
Abdulkareem Alali, Huzefa Kagdi, and Jonathan~I. Maletic. 2008.
\newblock \href {https://doi.org/10.1109/ICPC.2008.24} {What's a typical commit? a characterization of open source software repositories}.
\newblock In \emph{16th IEEE International Conference on Program Comprehension (ICPC)}, pages 182--191.

\bibitem[{Banerjee and Lavie(2005)}]{banerjee2005meteor}
Satanjeev Banerjee and Alon Lavie. 2005.
\newblock Meteor: An automatic metric for mt evaluation with improved correlation with human judgments.
\newblock In \emph{Proceedings of the ACL workshop on intrinsic and extrinsic evaluation measures for machine translation and/or summarization}, pages 65--72. Association for Computational Linguistics.

\bibitem[{Brown et~al.(2020)Brown, Mann, Ryder, Subbiah, Kaplan, Dhariwal, Neelakantan, Shyam, Sastry, Askell, Agarwal, Herbert-Voss, Krueger, Henighan, Child, Ramesh, Ziegler, Wu, Winter, Hesse, Chen, Sigler, Litwin, Gray, Chess, Clark, Berner, McCandlish, Radford, Sutskever, and Amodei}]{brown2020language}
Tom~B. Brown, Benjamin Mann, Nick Ryder, Melanie Subbiah, Jared Kaplan, Prafulla Dhariwal, Arvind Neelakantan, Pranav Shyam, Girish Sastry, Amanda Askell, Sandhini Agarwal, Ariel Herbert-Voss, Gretchen Krueger, Tom Henighan, Rewon Child, Aditya Ramesh, Daniel~M. Ziegler, Jeffrey Wu, Clemens Winter, and 12 others. 2020.
\newblock Language models are few-shot learners.
\newblock In \emph{Advances in Neural Information Processing Systems (NeurIPS)}, volume~33, pages 1877--1901.

\bibitem[{Daneshyan et~al.(2025)Daneshyan, He, Wu, and Zhou}]{daneshyan2025smartnote}
Farbod Daneshyan, Runzhi He, Jianyu Wu, and Minghui Zhou. 2025.
\newblock Smartnote: An llm-powered, personalised release note generator that just works.
\newblock \emph{arXiv preprint arXiv:2505.17977}.

\bibitem[{Dong et~al.(2022)Dong, Lou, Zhu, Sun, Li, Zhang, and Hao}]{dong2022fira}
Jinhao Dong, Yiling Lou, Qihao Zhu, Zeyu Sun, Zhilin Li, Wenjie Zhang, and Dan Hao. 2022.
\newblock Fira: fine-grained graph-based code change representation for automated commit message generation.
\newblock In \emph{Proceedings of the 44th International Conference on Software Engineering}, pages 970--981.

\bibitem[{Eliseeva et~al.(2023)Eliseeva, Sokolov, Bogomolov, Golubev, Dig, and Bryksin}]{eliseeva2023commit_chronicle}
Aleksandra Eliseeva, Yaroslav Sokolov, Egor Bogomolov, Yaroslav Golubev, Danny Dig, and Timofey Bryksin. 2023.
\newblock From commit message generation to history-aware commit message completion.
\newblock In \emph{2023 38th IEEE/ACM International Conference on Automated Software Engineering}, pages 723--735. IEEE.

\bibitem[{Fleiss(1971)}]{fleiss1971mns}
Joseph~L Fleiss. 1971.
\newblock Measuring nominal scale agreement among many raters.
\newblock \emph{Psychological bulletin}, 76(5):378.

\bibitem[{GitHub(2025)}]{github_api}
Inc. GitHub. 2025.
\newblock Github rest api v3 documentation.
\newblock \url{https://docs.github.com/en/rest}.
\newblock Accessed: 2025-05-31.

\bibitem[{Howard and Ruder(2018)}]{howard2018universal}
Jeremy Howard and Sebastian Ruder. 2018.
\newblock Universal language model fine-tuning for text classification.
\newblock In \emph{Proceedings of the 56th Annual Meeting of the Association for Computational Linguistics (ACL)}, pages 328--339.

\bibitem[{Hu et~al.(2016)Hu, Zhang, Bai, Yu, and Yang}]{hu2016influence}
Yan Hu, Jun Zhang, Xiaomei Bai, Shuo Yu, and Zhuo Yang. 2016.
\newblock \href {https://doi.org/10.1186/s40064-016-2897-7} {Influence analysis of github repositories}.
\newblock \emph{SpringerPlus}, 5(1):1268.

\bibitem[{Jiang et~al.(2021)Jiang, Zhu, Yang, Liang, and Zuo}]{jiang2021deeprelease}
Huaxi Jiang, Jie Zhu, Li~Yang, Geng Liang, and Chun Zuo. 2021.
\newblock Deeprelease: Language-agnostic release notes generation from pull requests of open-source software.
\newblock In \emph{2021 28th Asia-Pacific Software Engineering Conference}, pages 101--110. IEEE.

\bibitem[{Jiang et~al.(2017)Jiang, Armaly, and McMillan}]{jiang2017automatically}
Siyuan Jiang, Ameer Armaly, and Collin McMillan. 2017.
\newblock Automatically generating commit messages from diffs using neural machine translation.
\newblock In \emph{2017 32nd IEEE/ACM International Conference on Automated Software Engineering (ASE)}, pages 135--146. IEEE.

\bibitem[{Jiang et~al.(2023)Jiang, Goyal, Scialom, Muti, Qian, Simig, Webson et~al.}]{jiang2023mistral}
Yinhan Jiang, Naman Goyal, Thomas Scialom, Andrea Muti, Yichong Qian, Daniel Simig, Albert Webson, and 1 others. 2023.
\newblock Mistral 7b.
\newblock \url{https://mistral.ai/news/announcing-mistral-7b/}.

\bibitem[{Kamezawa et~al.(2022)Kamezawa, Nishida, Shimizu, Miyazaki, and Nakayama}]{kamezawa2022rnsum}
Hisashi Kamezawa, Noriki Nishida, Nobuyuki Shimizu, Takashi Miyazaki, and Hideki Nakayama. 2022.
\newblock Rnsum: A large-scale dataset for automatic release note generation via commit logs summarization.
\newblock In \emph{Proceedings of the 60th Annual Meeting of the Association for Computational Linguistics (Volume 1: Long Papers)}, pages 8718--8735.

\bibitem[{Klepper et~al.(2016)Klepper, Krusche, and Bruegge}]{klepper2016semi}
Sebastian Klepper, Stephan Krusche, and Bernd Bruegge. 2016.
\newblock Semi-automatic generation of audience-specific release notes.
\newblock In \emph{Proceedings of the International Workshop on Continuous Software Evolution and Delivery}, pages 19--22.

\bibitem[{Lewis(2019)}]{lewis2019bart}
M~Lewis. 2019.
\newblock Bart: Denoising sequence-to-sequence pre-training for natural language generation, translation, and comprehension.
\newblock \emph{arXiv preprint arXiv:1910.13461}.

\bibitem[{Lin(2004)}]{lin2004rouge}
Chin-Yew Lin. 2004.
\newblock Rouge: A package for automatic evaluation of summaries.
\newblock In \emph{Text summarization branches out: Proceedings of the ACL-04 workshop}, pages 74--81. Association for Computational Linguistics.

\bibitem[{Liu and Lapata(2019)}]{liu2019text}
Yang Liu and Mirella Lapata. 2019.
\newblock Text summarization with pretrained encoders.
\newblock In \emph{Proceedings of the 2019 Conference on Empirical Methods in Natural Language Processing and the 9th International Joint Conference on Natural Language Processing (EMNLP-IJCNLP)}, pages 3721--3731.

\bibitem[{Maalej and Happel(2010)}]{maalej2010work}
Walid Maalej and Hendrik Happel. 2010.
\newblock Work descriptions: analysis of informal comments generated by developers.
\newblock In \emph{ICSM Workshops 2010}.
\newblock Discusses developer comments, informal "work descriptions" in repositories.

\bibitem[{Mihalcea and Tarau(2004)}]{mihalcea2004textrank}
Rada Mihalcea and Paul Tarau. 2004.
\newblock Textrank: Bringing order into text.
\newblock In \emph{Proceedings of the 2004 conference on empirical methods in natural language processing}, pages 404--411.

\bibitem[{Min et~al.(2022)Min, Lyu, Holtzman, Artetxe, Lewis, Hajishirzi, and Zettlemoyer}]{min2022rethinking}
Sewon Min, Xinxi Lyu, Ari Holtzman, Mikel Artetxe, Mike Lewis, Hannaneh Hajishirzi, and Luke Zettlemoyer. 2022.
\newblock \href {https://doi.org/10.18653/v1/2022.emnlp-main.759} {Rethinking the role of demonstrations: What makes in-context learning work?}
\newblock In \emph{Proceedings of the 2022 Conference on Empirical Methods in Natural Language Processing (EMNLP)}, pages 11048--11064, Abu Dhabi, United Arab Emirates. Association for Computational Linguistics.

\bibitem[{Moratanch and Chitrakala(2017)}]{moratanch2017extractive}
N~Moratanch and S~Chitrakala. 2017.
\newblock A survey on extractive text summarization.
\newblock In \emph{2017 international conference on computer, communication and signal processing (ICCCSP)}, pages 1--6. IEEE.

\bibitem[{Moreno et~al.(2016)Moreno, Bavota, Di~Penta, Oliveto, Marcus, and Canfora}]{moreno2016arena}
Laura Moreno, Gabriele Bavota, Massimiliano Di~Penta, Rocco Oliveto, Andrian Marcus, and Gerardo Canfora. 2016.
\newblock Arena: an approach for the automated generation of release notes.
\newblock \emph{IEEE Transactions on Software Engineering}, 43(2):106--127.

\bibitem[{Nath and Roy(2022)}]{nath2022towards}
Sristy~Sumana Nath and Banani Roy. 2022.
\newblock Towards automatically generating release notes using extractive summarization technique.
\newblock \emph{arXiv preprint arXiv:2204.05345}.

\bibitem[{Papineni et~al.(2002)Papineni, Roukos, Ward, and Zhu}]{papineni2002bleu}
Kishore Papineni, Salim Roukos, Todd Ward, and Wei-Jing Zhu. 2002.
\newblock Bleu: a method for automatic evaluation of machine translation.
\newblock In \emph{Proceedings of the 40th Annual Meeting on Association for Computational Linguistics}, pages 311--318. Association for Computational Linguistics.

\bibitem[{Pennington et~al.(2014)Pennington, Socher, and Manning}]{pennington2014glove}
Jeffrey Pennington, Richard Socher, and Christopher~D Manning. 2014.
\newblock Glove: Global vectors for word representation.
\newblock In \emph{Proceedings of the 2014 conference on empirical methods in natural language processing (EMNLP)}, pages 1532--1543.

\bibitem[{Qian et~al.(2024)Qian, Xie, Han, Zhu, Liu, Chen et~al.}]{baichuan2024qwen2}
Yujia Qian, Yichong Xie, Xu~Han, Yitao Zhu, Shaohan Liu, Weizhu Chen, and 1 others. 2024.
\newblock \href {https://arxiv.org/abs/2405.10050} {Qwen2: The next generation of qwen models}.
\newblock \emph{Preprint}, arXiv:2405.10050.

\bibitem[{Rae et~al.(2019)Rae, Potapenko, Jayakumar, and Lillicrap}]{rae2019compressive}
Jack~W Rae, Anna Potapenko, Siddhant~M Jayakumar, and Timothy~P Lillicrap. 2019.
\newblock Compressive transformers for long-range sequence modelling.
\newblock In \emph{International Conference on Learning Representations (ICLR)}.

\bibitem[{Raffel et~al.(2020)Raffel, Shazeer, Roberts, Lee, Narang, Matena, Zhou, Li, and Liu}]{raffel2020t5}
Colin Raffel, Noam Shazeer, Adam Roberts, Katherine Lee, Sharan Narang, Michael Matena, Yanqi Zhou, Wei Li, and Peter~J. Liu. 2020.
\newblock Exploring the limits of transfer learning with a unified text-to-text transformer.
\newblock \emph{Journal of Machine Learning Research}, 21:1--67.

\bibitem[{Scarselli et~al.(2008)Scarselli, Gori, Tsoi, Hagenbuchner, and Monfardini}]{scarselli2008gnn}
Franco Scarselli, Marco Gori, Ah~Chung Tsoi, Markus Hagenbuchner, and Gabriele Monfardini. 2008.
\newblock The graph neural network model.
\newblock \emph{IEEE Transactions on Neural Networks}, 20(1):61--80.

\bibitem[{Shi et~al.(2022)Shi, Wang, Tao, Du, Zhang, Han, Zhang, and Sun}]{shi2022race}
Ensheng Shi, Yanlin Wang, Wei Tao, Lun Du, Hongyu Zhang, Shi Han, Dongmei Zhang, and Hongbin Sun. 2022.
\newblock Race: Retrieval-augmented commit message generation.
\newblock \emph{arXiv preprint arXiv:2203.02700}.

\bibitem[{Shihab et~al.(2009)Shihab, Hassan et~al.}]{shihab2013mining}
Emad Shihab, Ahmed~E. Hassan, and 1 others. 2009.
\newblock Mining software repositories to analyze release notes and maintenance activities.
\newblock \emph{CLEI Electronic Journal}.
\newblock Covers release notes as documentation sources in software evolution studies.

\bibitem[{Spadini et~al.(2018)Spadini, Aniche, and Bacchelli}]{spadini2018PyDriller}
Davide Spadini, Maur{\'\i}cio Aniche, and Alberto Bacchelli. 2018.
\newblock Pydriller: Python framework for mining software repositories.
\newblock In \emph{Proceedings of the 2018 26th ACM Joint Meeting on European Software Engineering Conference and Symposium on the Foundations of Software Engineering}, pages 908--911.

\bibitem[{Touvron et~al.(2024)Touvron, Lavril, Izacard, Martinet, Lachaux, Chan, Fu et~al.}]{touvron2024llama3}
Hugo Touvron, Thibaut Lavril, Gautier Izacard, Xavier Martinet, Marie-Anne Lachaux, Junjie Chan, Jie Fu, and 1 others. 2024.
\newblock \href {https://arxiv.org/abs/2404.14219} {Llama 3: Open foundation and instruction-tuned language models}.
\newblock \emph{Preprint}, arXiv:2404.14219.

\bibitem[{Wang et~al.(2021)Wang, Xia, Lo, He, Wang, and Grundy}]{wang2021context}
Haoye Wang, Xin Xia, David Lo, Qiang He, Xinyu Wang, and John Grundy. 2021.
\newblock Context-aware retrieval-based deep commit message generation.
\newblock \emph{ACM Transactions on Software Engineering and Methodology}, 30(4):1--30.

\bibitem[{Wu et~al.(2022)Wu, He, Xiao, Gao, and Zhou}]{wu2022demystifying}
Jianyu Wu, Hao He, Wenxin Xiao, Kai Gao, and Minghui Zhou. 2022.
\newblock Demystifying software release note issues on github.
\newblock In \emph{Proceedings of the 30th IEEE/ACM International Conference on Program Comprehension}, pages 602--613.

\bibitem[{Xu et~al.(2019)Xu, Yao, Xu, Gu, Tong, and Lu}]{xu2019commit}
Shengbin Xu, Yuan Yao, Feng Xu, Tianxiao Gu, Hanghang Tong, and Jian Lu. 2019.
\newblock Commit message generation for source code changes.
\newblock In \emph{IJCAI}.

\bibitem[{Yu(2009)}]{yu2009mining}
Liguo Yu. 2009.
\newblock Mining change logs and release notes to understand software maintenance and evolution.
\newblock \emph{CLEI Electronic Journal}, 12(2):1--10.

\end{thebibliography}
\clearpage
\appendix




\section{Dataset Details\label{app:data_collect}}
\label{app:dataset_detail}
Table~\ref{tab:attributes} summarizes all dataset attributes and their descriptions.  Compared to existing datasets, \textsc{ReleaseEval} provides more comprehensive information by including not only release notes and commit messages, but also commit metadata, commit trees, and fine-grained code diffs. Table~\ref{tab:tree_structure} illustrates the common symbols used in commit trees, including unique commits, branch divergence, and merge points.

\begin{table*}[htbp]
  \centering
  \small
  \begin{minipage}{0.95\textwidth}

    \centering
    \begin{tabular}{@{}ll@{}}
      \toprule
      \textbf{Attribute} & \textbf{Description} \\
      \midrule
      \multicolumn{2}{c}{\textit{Release information}} \\
      \midrule
      repo          & The URL of the repository. \\
      version\_url  & The URL linking to the specific release version in the repository. \\
      release\_note & The content of the release note. \\
      \midrule
      \multicolumn{2}{c}{\textit{Commits information}} \\
      \midrule
      commit\_url   & The URL linking to the comparison of commits between versions. \\
      hash          & The unique identifier (hash) of the commit. \\
      message       & A brief description of the changes introduced by the commit. \\
      author        & The person who authored the commit. \\
      time          & The date and time of the commit. \\
      commit\_tree  & The full commit tree between two versions (branches and merges). \\
      \midrule
      \multicolumn{2}{c}{\textit{Code diff information}} \\
      \midrule
      mod           & File modification type (added, modified, deleted). \\
      file          & Target modified files. \\
      diff          & The actual code changes introduced by the commit. \\
      \bottomrule
    \end{tabular}
    \caption{Basic information for each entry in \textsc{ReleaseEval}, including release information, commit information, and code diff information.}
    \label{tab:attributes}
  \end{minipage}
\end{table*}

\begin{table}
\centering
\small
\begin{tabularx}{\linewidth}{@{}lX@{}}
\toprule
\textbf{Symbol} & \textbf{Description} \\
\midrule
\texttt{*}      & Represents a unique commit. \\
\texttt{|\textbackslash} & Shows branch divergence (split or merge). \\
\texttt{|/}     & Marks a point where branches merge back. \\
\texttt{|*}     & Indicates a commit on a branch before it merges back. \\
\bottomrule
\end{tabularx}
\caption{Symbols in commit tree}
\label{tab:tree_structure}
\end{table}

\section{Experiments Details}
\label{app:exp_detail}

Table~\ref{tab:combined_config} summarizes the hyperparameters for both fine-tuning and few-shot settings, while Table~\ref{tab:combinedprompts} details the few‑shot prompts.

\section{Results Details}
\label{app:case_study}

Table~\ref{tab:compare_3_task_example} presents a case study comparing release notes from a fine-tuned Ministral-8B model on the three tasks (\textit{commit2sum}, \textit{tree2sum}, and \textit{diff2sum}). Colored highlights indicate ground-truth overlap, illustrating how input types affect coverage and quality.

\begin{table}
\centering
\small
\begin{tabularx}{\linewidth}{@{}lX@{}}
\toprule
\textbf{Category} & \textbf{Configuration} \\
\midrule
\multicolumn{2}{c}{\textit{Fine-Tuning}} \\
\midrule
Adaptation Method  & LoRA \\
Learning Rate      & $1\times10^{-4}$ \\
Batch Size         & 16 \\
Early Stopping     & Monitor validation loss (patience = 2) \\
Hardware           & 4× NVIDIA A100 GPUs \\
Repetitions        & 3 runs (results averaged) \\ 
\midrule
\multicolumn{2}{c}{\textit{Few-Shot}} \\
\midrule
Max Generation Length    & 256 tokens \\
Temperature        & 0.0 \\
Top-$p$            & 1.0 \\
Prompt Variants    & Zero-shot, two-shot, four-shot \\
\bottomrule
\end{tabularx}
\caption{Configuration of fine-tuning and few-shot settings.}
\label{tab:combined_config}
\end{table}

\renewcommand{\arraystretch}{1.15} 
\begin{table}[htbp]
  \centering
  \small
  \begin{tabularx}{\linewidth}{@{}p{0.20\linewidth}X@{}}
    \toprule
    \textbf{Type} & \textbf{Prompt} \\
    \midrule
    Zero-shot & Based on the given \texttt{<X>}, generate a release note summarizing 
                the key changes, including bug fixes, new features, enhancements, 
                and other modifications. \\
    \addlinespace
    Few-shot  & Based on the given \texttt{<X>}, generate a release note summarizing 
                the key changes, including bug fixes, new features, enhancements, 
                and other modifications. \newline
                I will provide you with \texttt{<N>} example(s): \newline
                [{\textit{Exemplars}}]
\\
    \bottomrule
  \end{tabularx}
  \caption{Prompts for LLMs, where \texttt{<X>} is the commit message, commit tree, 
           or code difference, and \texttt{<N>} is the total number of examples.}
  \label{tab:combinedprompts}
\end{table}
\begin{table*}[htbp]
\centering
\small
\setlength{\tabcolsep}{4pt}
\begin{tabularx}{\textwidth}{@{}lXc@{}}
\toprule
\textbf{Task} & \textbf{Release Notes} & \textbf{BLEU-4} \\
\midrule
Ground Truth &
\hlblue{What's Changed} \newline
\hspace*{1em}\hlblue{fix{:} prevent all referenced properties from updating to latest} regardless of recipe by \texttt{@ajohnsonz} \newline
\hspace*{1em}\hlblue{Add headers for FindCommitters data tables} by \texttt{@pstreef} \newline
\hspace*{1em}\hlblue{feat{:} add from date option to FindCommitters recipe} by \texttt{@pstreef} \newline
\hspace*{1em}\hlblue{Handle empty string option} by \texttt{@pstreef} \newline
\hspace*{1em}\hlblue{Make DependencyUseStringNotation compatible with G.GString} by \texttt{@Jenson3210} \newline
\hspace*{1em}\hlblue{Normalization of maven repositories that don't permit directory listing} by \texttt{@jkschneider} \newline
\hspace*{1em}\hlblue{New Contributors} \texttt{@Jenson3210} \hlblue{made their first contribution} \newline
\hspace*{1em}\hlblue{Full Changelog}: v8.14.3\ldots v8.14.4 & \\
\hdashline
commit2sum &
\hlorange{What's Changed} \newline
\hspace*{1em}Fix{:} \hlorange{prevent all referenced properties from updating to latest} regardless of recipe by \texttt{@timtebeek} \newline
\hspace*{1em} \hlorange{Add headers for FindCommitters data tables} by \texttt{@eduardoramirez} \newline
\hspace*{1em}Fix UpgradeDependencyVersion so it doesn't use the same GradleProject marker for every project in a multi-\hspace*{1em}project build. by \texttt{@timtebeek} \newline
\hspace*{1em}Add logic to simplify J.Ternary by \texttt{@jkschneider} \newline
\hspace*{1em}\hlorange{feat{:} add from date option to FindCommitters recipe} by \texttt{@eduardoramirez} \newline
\hspace*{1em}\hlorange{Handle empty string option} by \texttt{@eduardoramirez} \newline
\hspace*{1em}Add recipe which encloses all parameters interpolated into Groovy GStrings in curly braces by \texttt{@timtebeek} \newline
\hspace*{1em}Add FindClassHierarchy recipe to produce a data table of class declarations, their superclasses, and interfaces \hspace*{1em}by \texttt{@timtebeek} \newline
\hspace*{1em}\hlorange{Make DependencyUseStringNotation compatible with G.GString} by \texttt{@JenteSondervorst} \newline
\hspace*{1em}Improve AddProperty to insert in alphabetical order by key by \texttt{@timtebeek} \newline
\hspace*{1em}\hlorange{Normalization of maven repositories that don't permit directory listing} by \texttt{@sambsnyd} \newline
\hspace*{1em}\hlorange{New Contributors} \texttt{@JenteSondervorst} \hlorange{made their first contribution} \newline
\hspace*{1em}\hlorange{Full Changelog}: v8.23.2\ldots v8.23.3 & 39.95 \\
\midrule
tree2sum &
\hlgreen{What's Changed} \newline
\hspace*{1em}\hlgreen{fix{:} prevent all referenced properties from updating to latest} regardless of recipe by \texttt{@timtebeek} \newline
\hspace*{1em}\hlgreen{Add headers for FindCommitters data tables} by \texttt{@knutwannheden} \newline
\hspace*{1em}\hlgreen{feat{:} add from date option to FindCommitters recipe} by \texttt{@josh-farrant} \newline
\hspace*{1em}\hlgreen{Handle empty string option} by \texttt{@knutwannheden} \newline
\hspace*{1em}\hlgreen{Make DependencyUseStringNotation compatible with G.GString} by \texttt{@timtebeek} \newline
\hspace*{1em}Improve AddProperty to insert in alphabetical order by key by \texttt{@timtebeek} \newline
\hspace*{1em}\hlgreen{Normalization of maven repositories that don't permit directory listing} by \texttt{@knutwannheden} \newline
\hspace*{1em}\hlgreen{New Contributors} \texttt{@josh-farrant} \hlgreen{made their first contribution} \newline
\hspace*{1em}\hlgreen{Full Changelog}: v8.7.1\ldots v8.7.2 & \textbf{66.28} \\
\midrule
diff2sum &
\hlpink{What's Changed} \newline
\hspace*{1em}\hlpink{Add from date option to findCommitters} by \texttt{@timtebeek} \newline
\hspace*{1em}Add support for gradle plugin version notation by \texttt{@timtebeek} \newline
\hspace*{1em}Fix SimplifyBooleanExpressionVisitor by \texttt{@timtebeek} \newline
\hspace*{1em}\hlpink{New Contributors} \newline
\hspace*{1em}... \texttt{@tobiasnolte} \hlpink{made their first contribution} \newline
\hspace*{1em}\hlpink{Full Changelog}: v8.3.2\ldots v8.3.3 & 16.10 \\
\bottomrule
\end{tabularx}
\caption{Example release notes generated by a fine-tuned Ministral-8B on ReleaseEval across three tasks. Colors indicate overlap with the ground truth (blue),\textit{commit2sum} (orange), \textit{tree2sum} (green), and \textit{diff2sum} (pink).}
\label{tab:compare_3_task_example}
\end{table*}

\end{document}